\documentclass[aps,prb,twocolumn,superscriptaddress,longbibliography,reprint,floatfix]{revtex4-2}

\usepackage{graphicx}
\usepackage{comment}
\DeclareUnicodeCharacter{2212}{\uniminus}

\usepackage{gensymb}
\usepackage{float}
\usepackage{subfiles}
\usepackage{mathtools}  
\usepackage{amssymb}    
\usepackage{dirtytalk}
\usepackage{braket}
\usepackage{siunitx}
\usepackage{tikz}

\usepackage{stackengine}
\stackMath

\newcommand{\opbase}[1]{%
  \mathop{}\!\stackon[-.95ex]{#1}{\smash{\hat{}}}%
}

\newcommand{\op}[1]{\opbase{#1}^{\vphantom{\dagger}}}
\newcommand{\opdag}[1]{\opbase{#1}^{\dagger}}

\setcitestyle{super}

\begin{document}


\title{Collective Excitonic Entanglement in Singlet Fission}

\author{Lillian I. Payne}

\author{David A. Mazziotti}
\email{damazz@uchicago.edu}

\affiliation{Department of Chemistry and The James Franck Institute, The University of Chicago, Chicago, IL 60637}%

\date{Submitted July 31, 2026}


\begin{abstract}
Singlet fission, the conversion of one singlet exciton into two triplet excitons, has the potential to revolutionize solar energy production by boosting the efficiency of solar cells beyond the Shockley–Queisser limit. However, the practical realization of singlet fission remains an outstanding challenge, and its underlying mechanisms have long been debated. Here we show that singlet fission arises from a distinctive type of collective excitonic entanglement in which the excitons condense into a single particle-hole mode, the same principle that leads to the Bose-Einstein condensation of excitons. Quantitatively, this condensation appears as a large eigenvalue of the particle-hole reduced density matrix (RDM), representing multiple excitons occupying a collective mode. We utilize the particle-hole RDM to capture the intrinsic particle-hole correlations in singlet fission materials including acene crystals and covalently linked pentacene dimers, revealing how structural changes control singlet fission mechanisms and modulate the contributions of charge transfer states. The same characteristic enhancement of excitonic population in a single mode is common to the onset of exciton condensation and the mechanism of photosynthetic energy transfer. Together, our results establish a unifying framework for understanding the origins of singlet fission in terms of collective excitonic entanglement.
\end{abstract}

\maketitle

\enlargethispage{2\baselineskip}

\section{Introduction}
The efficiency of solar cells is fundamentally limited by recombination pathways that lead to loss of charge carriers.\cite{hannaSolarConversionEfficiency2006} Singlet fission (SF), the generation of two triplet excitons from a single photoexcited singlet exciton, can improve on these fundamental limits by generating more charge carriers for the same number of incoming photons.\cite{raoHarnessingSingletExciton2017,smithSingletFission2010} Singlet fission has also generated interest as a potential platform for optical switches or logic gates,\cite{boReversibleGatingSinglet2025,ullrichUnconventionalSingletFission2021} which has exciting implications for the emergent fields of quantum technologies. A number of singlet fission materials have been identified, including molecular crystals,\cite{singhLaserGenerationExcitons1965,burdettExcitedStateDynamics2010,thorsmolleMorphologyEffectivelyControls2009,congreveExternalQuantumEfficiency2013} aggregate materials,\cite{barfordSingletFissionLycopene2023,musserNatureSingletExciton2015,martinez-martinezPolaritonAssistedSingletFission2018} and chromophore dimers.\cite{boReversibleGatingSinglet2025,quardokusThroughBondThroughSpaceCoupling2012,johnsonHighTripletYield2010} Despite extensive experimental and theoretical investigation, the microscopic mechanism of singlet fission and the nature of the participating states remain incompletely understood. In particular, the coupling between the initial singlet exciton and the triplet-pair state, as well as the pathway by which the former evolves into the latter, remain active areas of investigation. Although numerous theoretical studies have examined the fission mechanism, the character of the triplet pair, and the possible role of charge-transfer states, most have relied on simplified model Hamiltonians or electronic-structure methods, such as conventional density functional theory (DFT), that limit a first-principles treatment of strong electron correlation. Consequently, the role of strong correlation in singlet fission remains poorly understood.

In this work we propose that singlet fission operates on the concentration of particle-hole entanglement into a single collective mode, the same principle that leads to exciton condensation,\cite{safaeiQuantumSignatureExciton2018,sagerBeginningsExcitonCondensation2022,schoutenExcitonCondensationMolecularScale2021,paynetorresMolecularOriginsExciton2024a} highly efficient photosynthetic energy transfer,\cite{schoutenExcitonCondensateLikeAmplificationEnergy2023,schoutenExcitonCondensateLikeEnergyTransport2025} and collective-entanglement enhanced quantum sensing.\cite{torresEntanglementWitnessesCondensation2025} This type of entanglement is witnessed by the appearance of a large eigenvalue from the particle-hole reduced density matrix (RDM), revealing a form of off-diagonal long-range order (ODLRO),\cite{garrodParticleHoleMatrixIts1969,hanamuraCondensationEffectsExcitons1977} albeit realized here in a finite system. Specifically, we connect the correlated triplet pair state in a singlet fission material to the emergence of a large triplet eigenvalue in the particle-hole RDM, which indicates multiple triplet excitons in the same coherent state.

We model a variety of singlet fission materials, including patches of acene molecular crystals as well as molecular chromophore dimers, with variational 2-RDM theory in order to capture strong electronic correlations. We find that a large triplet eigenvalue signature occurs in each singlet fission material, while singlet eigenvalues do not exceed one. Examining the exciton density corresponding to the singlet exciton and the entangled triplet pair reveals that both may have charge-transfer character, the extent of which is system dependent. Increasing the patch size modeled for the molecular solids also reveals that exciton density can be delocalized between three or more acene molecules, suggesting that the two-molecule fragments often used in theoretical modeling of singlet fission may be insufficient. Together these results put forth an ``excitonic entanglement" view of singlet fission that highlights the importance of accounting for strong correlation in modeling singlet fission.

\section{Theoretical Background}

The most basic---and historically prevalent---definition of singlet fission is a four-electron process involving two molecules. In this view, photoexcitation generates a singlet exciton (the S$_1$ state) localized on one of the molecules in the dimer. This exciton interacts with the adjacent S$_0$ molecule and produces a correlated pair of triplet excitons $^1$(TT) either directly or via some kind of mediated multi-step mechanism. Specifically, $^1$(TT) is generated via an electron hopping process that is either (1) concerted or (2) stepwise with the stepwise mechanism involving a charge transfer (CT) intermediate where the excitonic electron and hole are localized on different molecules.\cite{casanovaTheoreticalModelingSinglet2018, greysonMaximizingSingletFission2010}

Many theoretical studies have operated within this picture, constructing approximate diabatic bases utilizing single configurations of localized excitations from the frontier molecular orbitals of the system of interest.\cite{miyataTripletPairStates2019} Studies utilizing this four electron picture rely on the assumptions that (a) photoexcitation results in a singlet exciton which is (b) localized on one chromophore which then transforms into two triplet excitons which are (c) also localized on the two respective chromophores. However, these assumptions do not reflect the reality of the electronic structure of singlet fission materials such as molecular acene crystals, which feature highly delocalized electronic states.\cite{miyataTripletPairStates2019, monahanChargeTransferMediated2015}

To focus on the nature of the excitonic states in singlet fission from a correlated perspective while avoiding the above assumptions, we develop a theoretical methodology for studying singlet fission based on the particle-hole RDM. The particle-hole RDM describes all particle-hole pairs, or excitons, in a system, capturing the intrinsic particle-hole correlations. The term particle-hole correlation in this context means that whenever a particular single-particle state $i$ is occupied, there is a high probability of a corresponding state $j$ being unoccupied. The particle-hole RDM thus serves as a type of particle-hole pair correlation function with off-diagonal coherences, allowing for the characterization of the excitonic states in singlet fission while accounting for electron correlation and excitonic entanglement effects that are neglected in single-reference, localized excitation pictures.

In second quantization, the particle-hole RDM is defined as
\begin{equation}
^2G^\mathrm{i,j}_\mathrm{k,l} = \bra{\Psi}   \opdag{a}_\mathrm{i}\op{a}_\mathrm{j}\opdag{a}_\mathrm{l}\op{a}_\mathrm{k} \ket{\Psi}
\label{eq:Gmat}
\end{equation}
\noindent where $\ket{\Psi}$ is the $N$-electron wavefunction, $\opdag{a}_\mathrm{i}$ is the fermionic creation operator that creates an electron in orbital $i$, and $\op{a}_\mathrm{i}$ is the fermionic annihilation operator that destroys an electron in orbital $i$.\cite{safaeiQuantumSignatureExciton2018,garrodParticleHoleMatrixIts1969} In the context of singlet fission, we examine the particle-hole RDM for the ground state wavefunction $\ket{\Psi_\mathrm{g}}$ of a singlet fission material. Retaining only the particle-hole states $\ket{\Psi_\mathrm{s}}$ created by excitations from $\ket{\Psi_\mathrm{g}}$, we expand the expression for the matrix elements as
\begin{equation}
{}^2\Tilde{G}^\mathrm{i,j}_\mathrm{k,l}=\sum_\mathrm{s} \bra{\Psi_\mathrm{g}}   \opdag{a}_\mathrm{i}\hat{a}^{}_\mathrm{j} \ket{\Psi_\mathrm{s}}\bra{\Psi_\mathrm{s}}\opdag{a}_\mathrm{l}\hat{a}^{}_\mathrm{k} \ket{\Psi_\mathrm{g}}.
\label{eq:modGmat}
\end{equation}
\noindent This matrix thus spans the particle-hole pairs in the system such that diagonal elements correspond to the populations of particle-hole states and off-diagonal elements to the coherences between them. The eigenvalues of this modified particle-hole RDM are calculated as
\begin{equation}
^2\Tilde{G}\nu_\mathrm{i} = \lambda_\mathrm{i} \nu_\mathrm{i},
\label{eq:GmatEigEg}
\end{equation}
\noindent where the eigenvectors $\nu_\mathrm{i}$ of the modified particle-hole RDM define natural particle-hole states, while the corresponding eigenvalues $\lambda_\mathrm{i}$ quantify the strength or occupation of these states (modes) in the particle-hole correlations of the ground state. Importantly, the eigenvectors are particle-hole pair wavefunctions rather than wavefunctions of particular $N$-electron excited states.  In the absence of coherences between particle-hole states, the eigenvalues will be either zero or one.  However, correlated excitonic states such as the triplet pair state $^1$(TT) in singlet fission will manifest themselves as $\lambda > 1$, corresponding to greater-than-one occupancy of a single coherent particle-hole state.

\begin{figure}
    \centering
    \includegraphics[width=\linewidth, keepaspectratio]{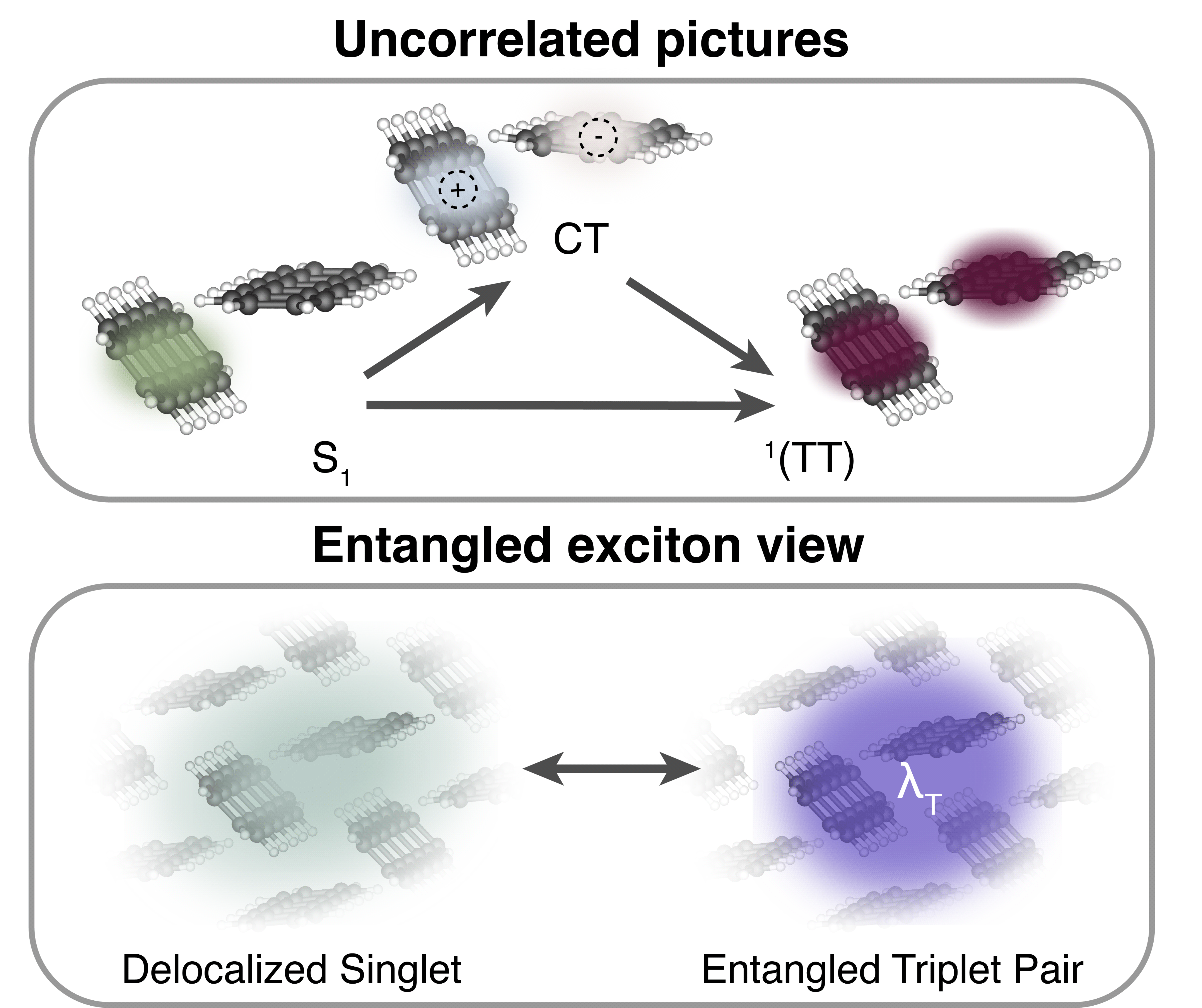}
    \caption{ Schematic comparing the traditional mechanisms proposed for singlet fission---either direct or mediated by a CT state (real or virtual)---to the entangled electron view presented here.  }
    \label{fig:theory_mech_compare}
\end{figure}

To understand how such a state emerges, consider the expectation value of the particle-hole RDM in Eq.~(\ref{eq:modGmat}):
\begin{align}
 \lambda &= \sum_\mathrm{ijkl} \nu^\mathrm{i}_\mathrm{j} {}^2\Tilde{G}^\mathrm{i,j}_\mathrm{k,l} \nu^\mathrm{l^*}_\mathrm{k} \\
& =\sum_\mathrm{s}\sum_\mathrm{ijkl} \nu^\mathrm{i}_\mathrm{j}
\nu^\mathrm{k^*}_\mathrm{l} \bra{\Psi_\mathrm{g}}   \opdag{a}_\mathrm{i}\hat{a}^\mathrm{}_\mathrm{j}
\ket{\Psi_\mathrm{s}}\bra{\Psi_\mathrm{s}}\opdag{a}_\mathrm{l}\hat{a}^\mathrm{}_\mathrm{k} \ket{\Psi_\mathrm{g}} \\
& =\sum_\mathrm{s} \bra{\Psi_\mathrm{g}}   \op{g} \ket{\Psi_\mathrm{s}}\bra{\Psi_\mathrm{s}}\opdag{g} \ket{\Psi_\mathrm{g}} ,
\end{align}
\noindent where the eigenoperator $\op{g}=\sum_\mathrm{i,j}^{}\nu_\mathrm{i,j}\opdag{a}_\mathrm{i}\op{a}_\mathrm{j}$ is constructed from the elements of the eigenvector $\nu_\mathrm{i,j}$ belonging to $\lambda$. In the absence of particle-hole pair correlations, $\op{g}$ corresponds to an incoherent transition and the eigenvalue cannot exceed one. However, if $\op{g}$ corresponds to a phase-coherent superposition of excitations, the corresponding eigenvalue can exceed one, indicating simultaneous occupation of a collective particle-hole state. The presence of this large eigenvalue is also related to the enhancement of transition amplitudes into such collective particle-hole states as long as the following two conditions are satisfied: (1) the constructed particle-hole state must have large overlap with an actual excited state of the system $\ket{\Psi_\mathrm{e}}\approx \op{g}\ket{\Psi_\mathrm{g}}$ and (2) the constructed transition operator $\op{g}$ must overlap strongly with an operator $\op{T}$ for a relevant physical process such as the dipole operator of an optical transition. If these conditions are met, the transition amplitude is defined as
\begin{equation}
\bra{\Psi_\mathrm{e}}   \op{T} \ket{\Psi_\mathrm{g}} \approx \lambda^\mathrm{\frac{1}{2}},
\end{equation}
\noindent indicating that it will be greatly enhanced as $\lambda$ grows large.

We can thus build an understanding of singlet fission in terms of the particle-hole RDM. For an ideal singlet-fission material, the expected value of the largest eigenvalue $\lambda_T$ in the triplet--triplet block is two, corresponding to two triplet excitons coherently occupying a collective quantum state. The associated eigenvector is the particle-hole mode associated with $^1$(TT), describing the particle-hole excitation from the singlet state to the coherently coupled triplet pair state in an inherently correlated fashion.  Thus, the emergence of a large eigenvalue in this block provides a direct signature of the collective triplet-pair character central to singlet fission.

Furthermore, as we will demonstrate, large triplet eigenvalues appear in the \textit{ground state} particle-hole RDMs of singlet fission materials.  These eigenvalues reveal intrinsic particle-hole correlations within the singlet ground state that are associated with an entangled collective state of triplet excitons. Although such a ground-state signature does not by itself establish the subsequent dynamics, it indicates the presence of a strongly favored coherent excitation channel connecting the singlet state to $^{1}(\mathrm{TT})$. This observation suggests that, rather than the triplet-pair state being generated only through an incoherent process following initial excitation into S$_1$, singlet fission may proceed via a coherent mechanism. This interpretation is supported by recent experiments demonstrating direct photoexcitation of the entangled triplet pair state.\cite{kimCoherentPhotoexcitationEntangled2024}

In the macroscopic limit, a large particle-hole eigenvalue indicates the emergence of macroscopic exciton condensation,\cite{safaeiQuantumSignatureExciton2018,garrodParticleHoleMatrixIts1969} an example of off-diagonal long-range order (ODLRO). This particle-hole signature of entanglement is not limited, however, to macroscopic exciton condensation. It has also been shown to characterize molecular-scale excitonic entanglement in photosynthetic light harvesting and quantum sensing.\cite{schoutenExcitonCondensateLikeAmplificationEnergy2023,schoutenExcitonCondensateLikeEnergyTransport2025,torresEntanglementWitnessesCondensation2025} Connecting this particle-hole entanglement signature with singlet fission therefore suggests an ``excitonic-entanglement'' picture that places singlet fission within a broader class of phenomena involving coherently correlated excitons.

\section{Results and Discussion}

\subsection{Singlet Fission in Acene Crystals}
\begin{figure}
    \centering
    \includegraphics[width=\linewidth, keepaspectratio]{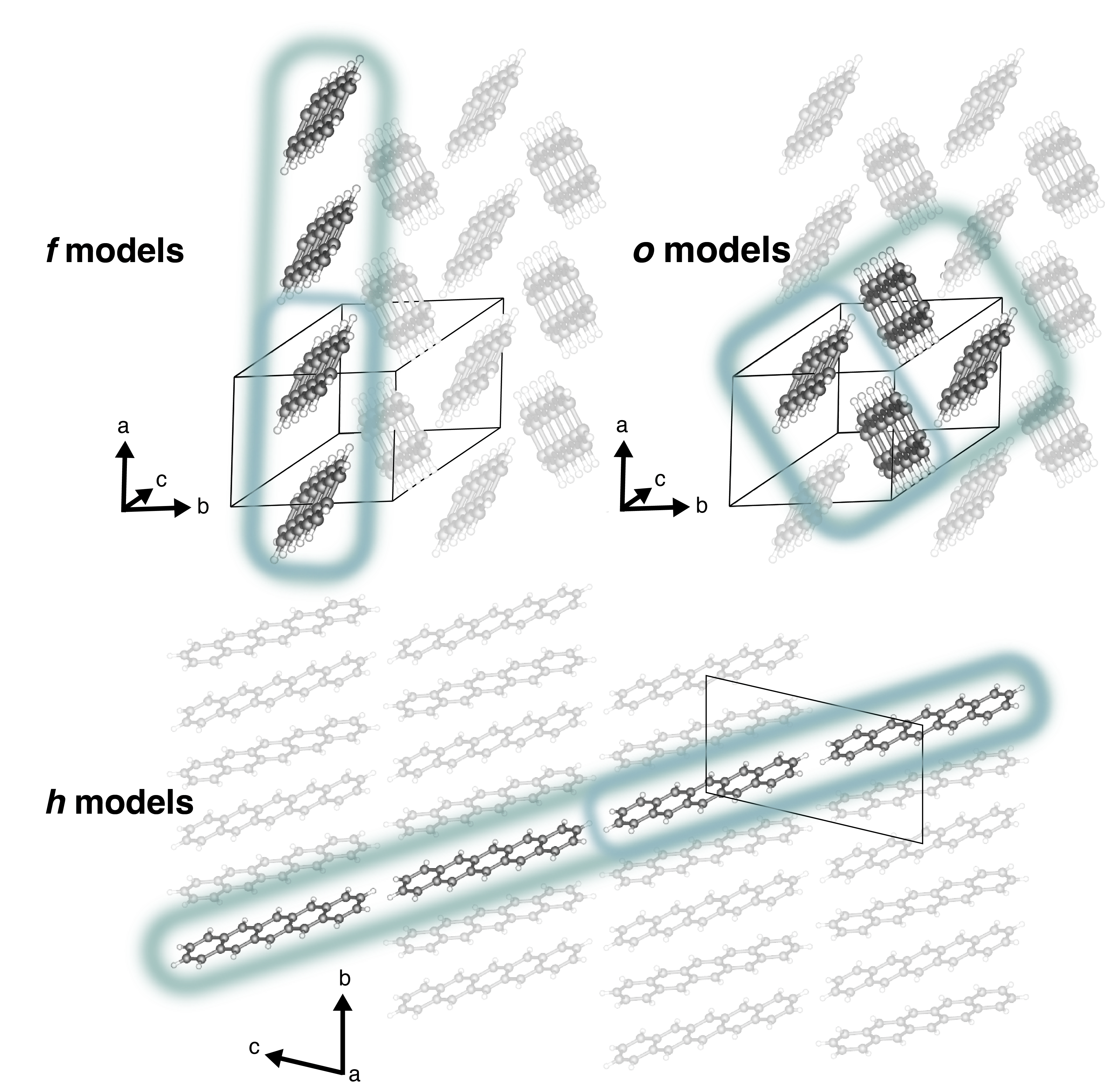}
    \caption{Model geometries for pentacene, demonstrating how each geometry is obtained by extracting a patch of molecules from the experimental crystal structure. Different geometric arrangements are considered for both the two molecule models (referred to as 2 unit in the following) and the four molecule models (referred to as 4 unit). These geometric arrangements can be grouped into three categories: pentacene molecules arranged along the crystallographic \textit{a} direction (labeled \textit{f}), molecules arranged in a cluster in the \textit{ab} plane (labeled \textit{o}) and molecules arranged end-to-end horizontally in the crystallographic \textit{c} direction (labeled \textit{h}).}
    \label{fig:pentacene_models}
\end{figure}
Acene-based molecular crystals such as crystalline pentacene are prototypical singlet fission materials that have been the subject of much experimental and theoretical study.\cite{singhLaserGenerationExcitons1965,burdettExcitedStateDynamics2010,zimmermanMechanismSingletFission2011,zimmermanSingletFissionPentacene2010,groffCoexistenceExcitonFission1970,geacintovEffectMagneticField1969,burgosHeterofissionPentacenedopedTetracene1977,thorsmolleMorphologyEffectivelyControls2009,thorsmollePhotoexcitedCarrierRelaxation2009} As such, we select three acene materials for our investigations: pentacene, tetracene, and perfluoropentacene. Singlet fission has traditionally been thought of as a dimeric process, and so the smallest model structure usable for modeling singlet fission in a crystalline solid would presumably consist of two acene molecules. However, experimental and theoretical studies have shown that the excitons in acene crystals may be highly delocalized, rendering such a simple model insufficient.\cite{berkelbachMicroscopicTheorySinglet2013, berkelbachMicroscopicTheorySinglet2014, chanEnergyBarrierSinglet2012} In light of this, we construct models consisting of two-and-four-molecule patches for each acene system. These patches are extracted from the experimental crystal structures. In addition, considering the effects of molecular orientations in the crystal, three different geometric arrangements are extracted for each model size. An example of these models is shown for pentacene in Fig.~\ref{fig:pentacene_models}. Models labeled \textit{f} feature molecules with face-on interactions, models labeled \textit{o} have molecules offset with face-to-edge interactions, and models labeled \textit{h} have molecules arranged horizontally end-to-end. The triplet and singlet exciton populations are computed for each model as described in the Theoretical Background section from the results of variational 2-electron reduced density matrix theory (v2RDM) calculations (for a more detailed description see Appendix~\ref{sec:pop}).
\begin{figure}
    \centering
    \includegraphics[width=\linewidth, keepaspectratio]{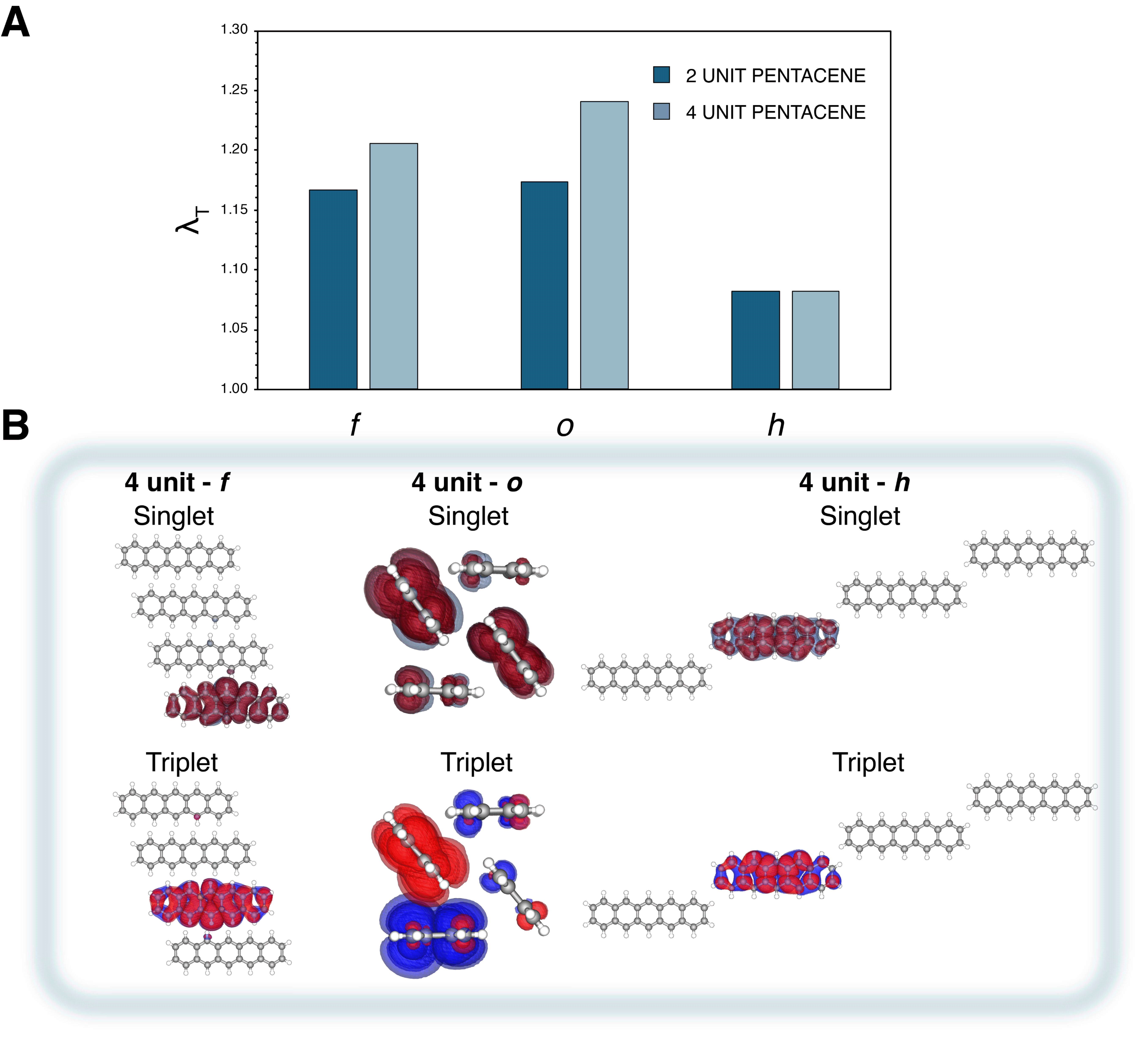}
    \caption{(A) Largest triplet eigenvalues $\lambda_T$ for each of the two and four unit pentacene models. The models are grouped into geometric arrangements labeled \textit{a}, \textit{o}, and \textit{h} as described in Fig.~\ref{fig:pentacene_models}, with the dark teal bars corresponding to the two unit models and the light teal bars corresponding to the four unit models. (B) Probabilistic exciton densities for the singlet (top) and triplet (bottom) excitons of the four unit models. The probabilistic density of the particle is plotted in red, while the probabilistic hole density is plotted in blue. }
    \label{fig:pentacene_2unit-4unit}
\end{figure}

Considering first the results for pentacene, we calculate the triplet and singlet eigenvalues for each of the 2- and 4-unit geometries. The largest triplet eigenvalue $\lambda_T$ exceeds one in each case, while the singlet eigenvalues do not exceed one. This indicates the possibility of more than one triplet exciton occupying the same coherent quantum state, while the singlet exciton states must be singly occupied. This is exactly what would be expected for singlet fission, which involves a \textit{two} coherently paired triplet excitons being produced from \textit{one} singlet exciton. The magnitude of $\lambda_T$ appears to depend on the size of the cluster model, as well as the geometric arrangement of pentacenes in the model. The \textit{f} and \textit{o} models, which feature face-face and face-edge interactions, have notably higher $\lambda_T$ values than the \textit{h} models, where molecules are arranged end-to-end. This suggests that face-on and face-edge interactions are important for singlet fission in pentacene. The \textit{f} and \textit{o} geometries also see an increase in $\lambda_T$ as the model size is increased from two to four molecules, supporting the idea that a simple two-molecule model is insufficient for capturing the full extent of singlet fission in pentacene. This increase is absent in the \textit{h} models, further suggesting that face-on and face-edge interactions drive singlet fission in pentacene.

While the calculated $\lambda_T$ values do not saturate the theoretical maximum value of two that might be expected for the case of the triplet pair state in singlet fission, this is not surprising given the limitations posed by a finite active space and model size. Comparing the $\lambda_T$ results from 2-unit calculations with an [12,12] (twelve electrons in twelve orbitals) active space as reported in Fig.~\ref{fig:pentacene_2unit-4unit} to those from calculations with a [18,18] active space reveals an increase in $\lambda_T$ for the larger active space. The results from these active space comparisons are provided in the Supplemental Information. These results, along with the fact that $\lambda_T$ increases between the two-unit and four-unit models, suggests that $\lambda_T$ may approach two as active space and model size increase.

To determine the nature of the exciton distribution for the triplet pair state, we visualize exciton density by calculating the probabilistic location of the hole corresponding to an electron localized in a specific atomic orbital (see Methods for more information). This visualization is performed for each of the pentacene model geometries, and the resulting plots are provided in Fig.~\ref{fig:pentacene_2unit-4unit}B. These plots reveal some degree of exciton delocalization in each case, with the delocalization strongest in the \textit{o} model. The exciton density in the four unit \textit{o} model is delocalized over all four molecules in the \textit{ab} plane, suggesting that the combination of face-on and face-edge interactions leads to significant delocalization. Comparing the density of the singlet exciton to the triplet pair state for the four-unit \textit{o} model (Fig.~\ref{fig:pentacene_2unit-4unit}B, center), we see significant delocalization for the singlet exciton as well. This aligns with previous theoretical work identifying the singlet exciton in pentacene as being highly delocalized with significant charge transfer (CT) contribution.\cite{sharifzadehLowEnergyChargeTransferExcitons2013} In this context, CT excitons are those with electron and hole density on different molecules. Together, the results here demonstrate that both the singlet exciton and the correlated triplet pair state in pentacene singlet fission have significant CT character.

\begin{figure}
    \centering
    \includegraphics[width=\linewidth, keepaspectratio]{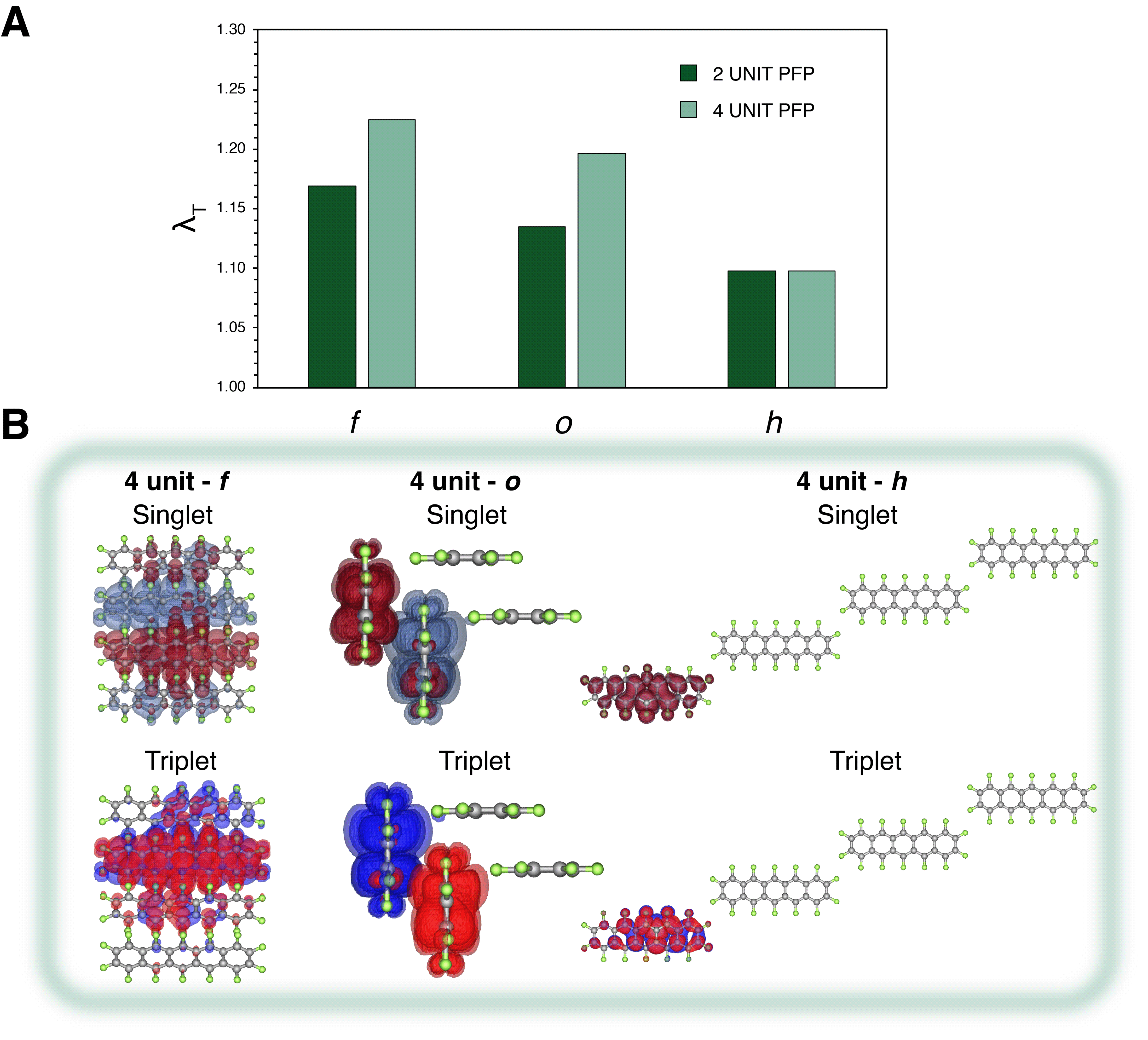}
    \caption{(A) Largest triplet eigenvalues $\lambda_T$ for each of the two and four unit perfluoropentacene (PFP) models. The models are grouped into geometric arrangements labeled \textit{b}, \textit{o}, and \textit{h} as described in Fig.~\ref{fig:pentacene_models}, with the dark green bars corresponding to the two unit models and the sage green bars corresponding to the four unit models. (B) Probabilistic exciton densities for the singlet (top) and triplet (bottom) excitons of the four unit models. The probabilistic density of the particle is plotted in red, while the probabilistic hole density is plotted in blue. }
    \label{fig:PFP_2unit-4unit}
\end{figure}

Perfluoropentacene (PFP, C$_{22}$F$_{14}$) is a singlet fission material wherein molecular packing has been shown to have a central impact on the singlet fission process.\cite{kolataMolecularPackingDetermines2014} The structure of crystalline PFP features anisotropic molecular packing in the \textit{bc} plane, with molecules slip stacked along the \textit{b} axis and aligned face-to-edge in a ``herringbone" arrangement along the \textit{c} axis. The extreme $90\degree$ angle between the molecules in the \textit{c} axis hinders interactions in that direction, and previous work has observed singlet fission in PFP occurring only along the \textit{b} axis. We perform calculations on models of PFP in the same manner as the calculations on pentacene in order to determine whether PFP also shows a signature of the correlated triplet pair state in the particle-hole RDM, and correlated modeling of the excitons produces the same packing-dependent effects observed previously.

Calculating the singlet and triplet eigenvalues for two-unit and three-unit models of each geometric arrangement for PFP once again produces a triplet eigenvalue $\lambda_T > 1$ for each case, while the singlet exciton states are predicted to be singly occupied. The value of $\lambda_T$ is maximized for the \textit{f} models, which are arranged slip-stacked along the crystallographic \textit{b} direction, and minimized for the \textit{h} models featuring molecules arranged end-to-end (Fig. \ref{fig:PFP_2unit-4unit}A. Plots of the triplet exciton density reveal that the triplet excitons delocalize along slip-stacked molecules in the \textit{b} direction, while minimal delocalization occurs between face-edge herringbone molecules along the \textit{c} axis (Fig.~\ref{fig:PFP_2unit-4unit}B). This supports the previous observations that singlet fission occurs primarily between PFP molecules along the \textit{b} axis due to stronger interactions between those molecules. Comparing the singlet exciton density to the exciton density of the triplet pair state for PFP (Fig.~\ref{fig:PFP_2unit-4unit}B, center) reveals that singlet exciton density delocalizes even further along the \textbf{b} direction, with the singlet hole delocalized across all four slip-stacked molecules in the four-unit \textit{f} model. Together, these results suggest that singlet fission in PFP is highly influenced by geometry-dependent molecular coupling, and involves triplet and singlet excitons with strong CT character.

\subsection{Singlet Fission in Molecular Dimers}
\begin{figure}
    \centering
    \includegraphics[width=\linewidth, keepaspectratio]{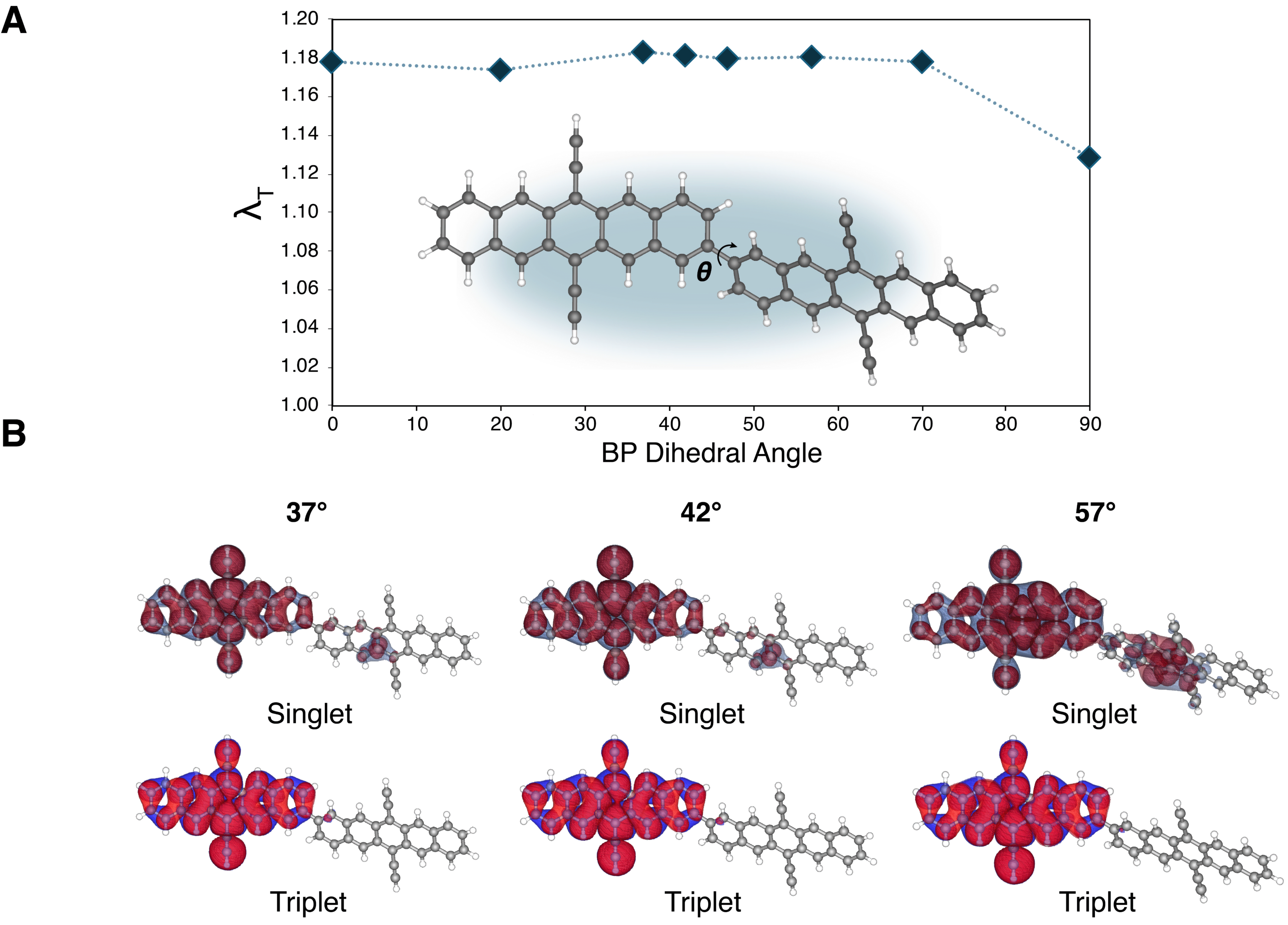}
    \caption{(A) Largest triplet eigenvalues $\lambda_T$ plotted as a function of bipentacene dihedral angle $\theta$. A representative schematic of the bipentacene molecule with the dihedral angle labeled is inset below the data points. (B) Probabilistic exciton densities for the singlet (top) and triplet (bottom) exciton modes for BP-37, BP-42, and BP-57. The probabilistic density corresponding to the particle is plotted in red, while the probabilistic hole density is plotted in blue. }
    \label{fig:BP-2H}
\end{figure}

Covalently coupled pentacene dimers have more recently become of interest for singlet fission materials due to the advantages they hold over traditional acene crystal materials, including highly tunable geometric and electronic structure and solution processability. These covalently coupled chromophores have been shown to participate in intramolecular singlet fission (iSF), wherein two triplet excitons are generated on one molecule consisting of covalently bonded acene moieties.\cite{zirzlmeierSingletFissionPentacene2015, fuemmelerDirectMechanismUltrafast2016, kimCoherentPhotoexcitationEntangled2024, sandersQuantitativeIntramolecularSinglet2015,  lukmanTuneableSingletExciton2015, busbyDesignStrategyIntramolecular2015, varnavskiHighYieldUltrafast2015, sandersIntramolecularSingletFission2016, sakumaLongLivedTripletExcited2016} As in the case of intermolecular singlet fission (xSF) in molecular crystals, there has been extensive research and debate over the mechanism of iSF and the potential involvement of CT states. To expand our investigation of the nature of the excitonic states and mechanisms in singlet fission to the case of iSF, we select two varieties of covalently linked pentacene dimers. The first is a directly 2,2'-connected bipentacene dimer referred to as BP-$\theta$ ($\theta$ being the dihedral angle) and the second is a 6,6'-connected dimer wherein the pentacenes are linked with a phenylene spacer in an \textit{ortho} (BP-\textit{o}), \textit{meta} (BP-\textit{m}), or \textit{para} (BP-\textit{p}) arrangement. For BP-$\theta$, eight models are constructed for dihedral angles from 0-90$\degree$. A representative structure for the BP-$\theta$ bipentacene molecules is provided in the inset of Fig. \ref{fig:BP-2H}A. For the phenylene-spaced bipentacene, models are constructed for each of the three isomers. The triplet and singlet exciton populations are once again calculated for each model as described in the Theoretical Background section from the results of variational 2-electron reduced density matrix theory (v2RDM) calculations (see Appendix~\ref{sec:v2rdm} for further details).

BP-37---the molecule depicted in Fig.~\ref{fig:BP-2H}A with a 37$\degree$ dihedral angle---is the parent molecule of the BP-$\theta$ family, and is a simplified version of the experimentally measured molecule BP-2H wherein the tri-isopropylsilyl groups of BP-2H are replaced with hydrogens. BP-2H has demonstrated singlet fission with an ~200 \% triplet yield,\cite{sandersQuantitativeIntramolecularSinglet2015} and the simplified BP-37 structure has been used in previous theoretical work to propose a direct mechanism for iSF in BP-2H and related molecules.\cite{fuemmelerDirectMechanismUltrafast2016} Calculating the singlet and triplet eigenvalues for each dihedral angle of BP-$\theta$, we find $\lambda_T > 1$ for each angle (Fig.~\ref{fig:BP-2H}A). In general $\lambda_T$ does not vary greatly with dihedral angle, although it is somewhat decreased for BP-90, suggesting a disruption to the coupling between the ground state and the triplet pair state.  Comparing the exciton densities for the triplet and singlet exciton modes in Fig.~\ref{fig:BP-2H}B supports previous findings of minimal CT involvement for iSF in this family of molecules---both singlet and triplet exciton densities are localized and lack significant CT character.

\begin{figure}
    \centering
    \includegraphics[width=\linewidth, keepaspectratio]{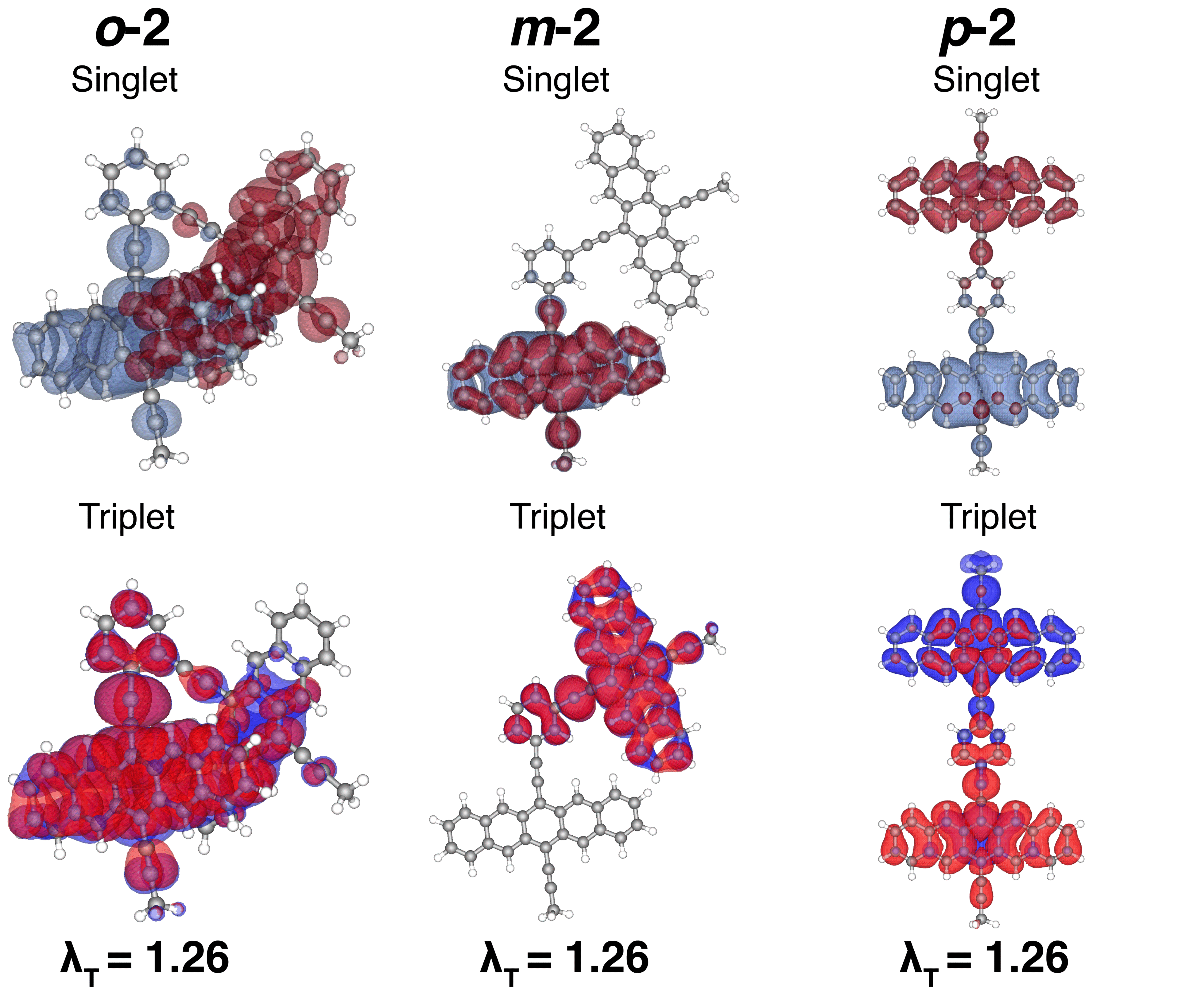}
    \caption{Probabilistic exciton densities for the exciton mode corresponding to $\lambda_T$ for the three phenylene-spaced pentacene dimer isomers. The atomic orbital corresponding to the constrained particle location is plotted in red, while the probabilistic hole density is plotted in purple. The value of $\lambda_T$ is provided below each exciton plot, and equals 1.26 in each case. }
    \label{fig:pentaquin}
\end{figure}

The phenylene-spaced bipentacene isomers BP-\textit{o}, BP-\textit{m}, and BP-\textit{p} have also experimentally demonstrated iSF, but unlike in the case of BP-$\theta$ evidence points towards a CT-mediated mechanism.\cite{zirzlmeierSingletFissionPentacene2015} Our calculations of the particle-hole eigenstates for the three isomers reveal equivalent values of $\lambda_T=1.26$ in each case. Comparing the exciton densities for the triplet and singlet exciton modes in Fig.~\ref{fig:pentaquin} reveals highly delocalized singlet and triplet excitons with significant charge transfer character for both BP-\textit{o} and BP-\textit{p}. This is in contrast to BP-\textit{m}, which features localized singlet and triplet excitons. However, the second singlet eigenstate of BP-\textit{m} has significant CT character, and contributions of this state to the iSF mechanism may explain experimental evidence that points toward a CT-involved mechanism for BP-\textit{m}.\cite{zirzlmeierSingletFissionPentacene2015}

In general, it appears that iSF can proceed from a variety of mechanisms, with the preferred mechanism strongly influenced by the connectivity-dependent inter-acene coupling in the dimers.\cite{kimCoherentPhotoexcitationEntangled2024,quardokusThroughBondThroughSpaceCoupling2012,zirzlmeierSingletFissionPentacene2015} Strong through-bond coupling in BP-\textit{p} and through-space coupling in BP-\textit{o} leads to highly delocalized singlet and triplet excitons with significant charge-transfer character. The meta isomer BP-\textit{m} features more localized singlet and triplet excitons, likely due to weakened through-bond interactions for the meta substitution pattern.\cite{quardokusThroughBondThroughSpaceCoupling2012} As previously observed,\cite{kimCoherentPhotoexcitationEntangled2024,fuemmelerDirectMechanismUltrafast2016} the dimers in BP-$\theta$ are weakly interacting, leading to primarily localized excitons without CT character. Together these results demonstrate that the ``excitonic entanglement'' view of singlet fission is appropriate both for xSF and iSF, and can be used to characterize the involvement of CT states in both xSF and iSF mechanisms. Furthermore, the emergence of a large triplet eigenvalue $\lambda_T$ in the ground state of iSF materials is connected to the favorability of a coherent superposition of excitations. This supports the possibility for iSF mechanisms involving coherent photoexcitation of the triplet pair state, as proposed by recent experimental work.\cite{kimCoherentPhotoexcitationEntangled2024}

\section{Conclusions and Outlook}

In this work we propose that singlet fission operates on the concentration of particle-hole entanglement into a single triplet collective mode, facilitating the favorable formation of the crucial triplet pair state in a microscopic version of the mechanism leading to exciton condensation. This type of entanglement is characterized by the appearance of a large eigenvalue from the particle-hole reduced density matrix (RDM), a theoretical entanglement witness first applied to exciton condensation,\cite{safaeiQuantumSignatureExciton2018, sagerBeginningsExcitonCondensation2022, schoutenExcitonCondensationMolecularScale2021, paynetorresMolecularOriginsExciton2024a} and later to highly efficient photosynthetic energy transfer,\cite{schoutenExcitonCondensateLikeAmplificationEnergy2023, schoutenExcitonCondensateLikeEnergyTransport2025} and collective-entanglement enhanced quantum sensing.\cite{torresEntanglementWitnessesCondensation2025} The triplet pair state in singlet fission can thus be thought of as possessing a microscopic form of off-diagonal long-range order (ODLRO), connecting it to a broader family of entangled excitonic states unified by this particle-hole entanglement witness.

We build a theoretical methodology for studying singlet fission based on the particle-hole reduced density matrix (RDM), which captures the intrinsic particle-hole correlations of a system that are neglected in the localized, single reference picture offered by single-electron theories. We construct the particle-hole RDMs for representative singlet fission materials---including acene molecular crystals as well as molecular chromophore dimers---from the results of variational 2-RDM theory, which captures strong electronic correlations. We find that large triplet eigenvalues appear in the particle-hole RDMs of each material, revealing intrinsic particle-hole correlations connecting the ground state to an entangled collective state of triplet excitons. The emergence of such a large eigenvalue is connected to the favorability of a coherent superposition of excitations, suggesting that singlet fission may proceed via a coherent mechanism, an idea supported by recent experiments demonstrating direct photoexcitation of the entangled triplet pair state.\cite{kimCoherentPhotoexcitationEntangled2024}

Characterizing the singlet and triplet excitons via exciton density calculated from the particle-hole RDM reveals that both may have charge-transfer (or charge-resonance) character, the extent of which is system dependent. Both the singlet and triplet excitons in the acene crystal models are delocalized with charge transfer character, and the entangled triplet pair appears highly sensitive to geometry-dependent molecular couplings. Structural changes such as dihedral angle or substitution in molecular dimers modulate the contributions of charge transfer states, supporting the idea that multiple singlet fission pathways can be available for the same material.

Together these results put forth an ``excitonic entanglement'' view of singlet fission that moves beyond the artificial distinctions between ``charge-transfer'' and ``direct'' mechanisms and highlights the importance of accounting for strong correlation in modeling singlet fission. The particle-hole entanglement witness offers a unifying lens that connects triplet pair formation in singlet fission to a family of related excitonic phenomena, which both deepens fundamental understandings of the process and may inspire future design principals for singlet fission materials.

\begin{acknowledgments}
D.M. gratefully acknowledges the U.S. Department of Energy, Office of Science, Basic Energy Sciences under Award No. DE-SC0026076 for support. This project is supported, in part, by funding from Jump Trading. Any opinions, findings, and
conclusions or recommendations expressed in this material are those of the authors and do not reflect
the views of Jump Trading.
\end{acknowledgments}

\appendix

\section{Electronic Structure Calculations}

\label{sec:v2rdm}

We employ variational two-electron reduced density matrix theory (v2RDM) to compute the 2-RDM.\cite{mazziottiReduceddensitymatrixMechanicsApplications2007,mazziottiContractedSchrodingerEquation1998,nakataVariationalCalculationsFermion2001,mazziottiRealizationQuantumChemistry2004,gidofalviActivespaceTwoelectronReduceddensitymatrix2008,mazziottiTwoElectronReducedDensity2012,mazziottiPureRepresentabilityConditions2016,mazziottiLargeScaleSemidefiniteProgramming2011,cioslowskiManyelectronDensitiesReduced2000,zhaoReducedDensityMatrix2004,cancesElectronicGroundstateEnergy2006,shenviActiveSpaceRepresentabilityConstraints2010,mazziottiStructureFermionicDensity2012,verstichelVariationalTwoParticleDensity2012,pirisGlobalMethodElectron2017, Knight.2022, Gao.2025dgs, Mazziotti.2026, Schouten.2026} V2RDM minimizes the energy as a functional of the 2-RDM subject to a set of \textit{N}-representability conditions which ensure correspondence between the 2-RDM and a physical \textit{N}-electron density matrix. Here we apply the basic conditions of Hermiticity, normalization, and antisymmetry, as well as additional constraints known as the 2-positivity conditions. These 2-positivity conditions, also known as the DQG conditions, constrain the two-particle ($^2D^\mathrm{i,j}_\mathrm{k,l} = \bra{\Psi}   \opdag{a}_\mathrm{i}\opdag{a}_\mathrm{j}\op{a}_\mathrm{l}\op{a}_\mathrm{k} \ket{\Psi}$), two-hole ($^2Q^\mathrm{i,j}_\mathrm{k,l} = \bra{\Psi}   \op{a}_\mathrm{i}\op{a}_\mathrm{j}\opdag{a}_\mathrm{l}\opdag{a}_\mathrm{k} \ket{\Psi} $), and particle-hole $^2G^\mathrm{i,j}_\mathrm{k,l} = \bra{\Psi} \op{a}_\mathrm{i}\opdag{a}_\mathrm{j}\opdag{a}_\mathrm{l}\op{a}_\mathrm{k} \ket{\Psi} $ RDMs to be positive semidefinite (A matrix is positive semidefinite if and only if its eigenvalues are nonnegative).  In this work we specifically utilize v2RDM within a complete-active-space self-consistent-field (CASSCF) calculation in which the 2-RDM calculation is applied to a set of active orbitals that is iteratively optimized to yield the lowest energy.\cite{gidofalviActivespaceTwoelectronReduceddensitymatrix2008}.  The semidefinite programming is solved with a boundary-point algorithm~\cite{mazziottiLargeScaleSemidefiniteProgramming2011}.  These calculations are performed in the Quantum Chemistry Toolbox for Maple.\cite{Maple-2026, Maple-QC-2026}

For the acene crystal systems, a [12,12] active space is employed for all 2-unit models while a [24,24] active space is employed for all 4-unit models, where X,Y] indicates an active space of X electrons in Y orbitals. For the molecular dimers, an [18,18] active space is used for each case. Trials of various basis sets reveal that basis set has minimal impact on the results, and hence, the a minimal Slater-type-orbital with 6 Gaussians (STO-6G) basis set is used for all calculations to minimize computational expense.\cite{hehreSelfConsistentMolecularOrbitalMethods1969}

From the results of the v2RDM calculation we construct the modified particle-hole RDM from the 2-RDM. The particle-hole RDM can be obtained from the 2-RDM by the following linear mapping
\begin{equation}
^2G^\mathrm{i,j}_\mathrm{k,l}={}^1I^\mathrm{j}_\mathrm{l} {}^{1}D^\mathrm{i}_\mathrm{k} - {}^{2}D^\mathrm{i,l}_\mathrm{k,j}
\label{eq:GfromD}
\end{equation}
\noindent where ${}^{2}D^\mathrm{i,l}_\mathrm{k,j}$ is the 2-RDM, $\delta^\mathrm{j}_\mathrm{l}$ is the identity matrix, and $^{1}D^\mathrm{i}_\mathrm{k}$ is the one-electron reduced density matrix (1-RDM).  The particle-hole RDM is spin adapted.  Diagonalization of the singlet and triplet blocks yield the singlet and triplet exciton eigenvalues and eigenvectors, respectively. The detailed structure of these blocks is as described in Ref. \cite{gidofalviSpinSymmetryAdaptation2005}. In this work, the largest singlet eigenvalue is referred to as $\lambda_S$ while the largest triplet eigenvalue is referred to as $\lambda_T$.

\section{Visualization of Exciton Density}

\label{sec:pop}

Visualization of the excitonic states corresponding to $\lambda_S$  and $\lambda_T$ are obtained by constraining the location of the excitonic electron to a particular atomic orbital and plotting the probabilistic density of the hole. To generate these plots, a matrix of the molecular orbitals in terms of atomic orbitals ($M_\mathrm{MO,AO}$) is extracted from the output of v2RDM calculations. This matrix is then used to calculate a matrix of atomic orbitals in terms of molecular orbitals, $M_\mathrm{AO,MO}$
\begin{equation}
M_\mathrm{AO,MO}=(M^\mathrm{T}_\mathrm{MO,AO})^{-1}.
\end{equation}
\noindent We isolate a submatrix $M^{\rm active}_\mathrm{AO,MO}$ corresponding to the active orbitals from $M_\mathrm{AO,MO}$, and the eigenvector of the modified particle-hole RDM corresponding to $\lambda_S$ or $\lambda_T$ is reshaped as a matrix in the basis of the active molecular orbitals. This matrix, called $V_{\rm max}$, is multiplied along with $M^\mathrm{active}_\mathrm{AO,MO}$ to create a matrix representing hole atomic orbitals in terms of the coefficients of the contributions of the electrons to other molecular orbitals (and vice versa)
\begin{equation}
(M^\mathrm{active}_\mathrm{AO,MO})(V_\mathrm{max})(M^\mathrm{active}_\mathrm{AO,MO})^\mathrm{T}.
\end{equation}
\noindent Visualization tools in Maple are then used to construct a visual representation of the probabilistic hole (electron) density corresponding to a constrained electron (hole) location using these coefficients for a given excitonic mode. The constrained locations of the electrons and holes are chosen by first selecting the active molecular orbitals (electron and hole) that contribute predominantly to $V_{\rm max}$. The electron and hole are then localized on the atomic orbital that contributes most significantly to their respective active molecular orbital, and the corresponding probabilistic densities are plotted. In this work we present the probabilistic electron and hole densities together in one image.

\bibliography{References}

\end{document}